# Giant magnetodielectric metamaterial

Ke Bi, Ji Zhou[*], Xiaoming Liu

*State Key Laboratory of New Ceramics and Fine Processing, School of Materials Science and Engineering, Tsinghua University, Beijing 100084, P.R. China*
*Correspondence and requests for materials should be addressed to J.Z. (email: zhouji@mail.tsinghua.edu.cn).*

Dielectric materials with tunable permittivity are highly desirable for wireless communication, radar technology. However, the tunability of dielectric properties in the microwave frequency range and higher is an immense challenge for conventional materials. Here, we demonstrate a giant magnetodielectric effect in the GHz region in a metamaterial based on ferrite unit cells. The effect is derived from the coupling of the ferromagnetic resonance and the Mie resonance in the ferrite unit cells. Both the simulated and experimental results indicate that the effective permittivity of the metamaterial can be tuned by modifying the applied magnetic field, and a giant magnetodielectric effect, $[\varepsilon'(H) - \varepsilon'(0)]/\varepsilon'(0) = 15000\ \%$ at 11.284 GHz, is obtained. This mechanism offers a promising means of constructing microwave dielectrics with large tunable ranges and considerable potential for tailoring via a metamaterial route.

With the progressing development of wireless communication and radar technology, microwave dielectrics with tunable permittivity are highly desired for use as key materials in phase shifters, switches, reconfigurable antenna, and other tunable devices[1]. However, it is an immense challenge to obtain dielectric materials with high tunability at microwave frequencies and higher. Electrically tunable ferroelectric dielectrics, such as barium strontium titanate (BST), are well-studied candidates for this purpose[2,3]. However, these materials have certain limitations, most notably a low tunability scale of less than 21.3 % and high driving power; and for these reason, only the dielectric thin films are available for potential applications[2]. The magnetodielectric effect (MD) offers another approach to permittivity tuning[4,5]. Chen et al.[6] have reported a giant MD effect ($\Delta\varepsilon'/\varepsilon'$=1800 % at 3.5 kOe) and a large magnetic-field-tunable dielectric resonance in spinel MnZn ferrite. However, the MD effect in those materials appears in very low frequency (MHz) region and cannot reach the GHz range. Castel and Brosseau[7] have reported an MD effect at 4.5 GHz in $BaTiO_3$–Ni nanocomposites. However, this MD effect ($\Delta\varepsilon'/\varepsilon'$=10 % at 2 kOe) is very small.

Metamaterials are a class of artificial materials in which subwavelength features, rather than the features of the constituent materials, control the macroscopic electromagnetic properties[8]; this allows more freedom in tailoring of material properties[9-12]. Recently, Mie-resonance-based dielectric metamaterials with unusual electromagnetic properties have been theoretically and experimentally studied[13-15]. The permittivity of a dielectric metamaterial is dependent only on its electromagnetic parameters ($\varepsilon$ and $\mu$), the geometry of the dielectric unit cell and the cell's lattice arrangement[16-18]. In the work, we demonstrate a giant MD effect in the GHz region in a dielectric metamaterial based on the coupling of the Mie resonance and ferromagnetic resonance of ferrite in the unit cells.

**Results**

In the metamaterial, ferrite rods are used as the unit cells to generate the Mie resonance, as illustrated in Fig. 1. Meanwhile, by interacting with the magnetic field of an electromagnetic wave, ferromagnetic resonance can arise in ferrite under an applied magnetic field. The effective permeability of the ferrite around the frequency area of ferromagnetic resonance can be expressed as follows[21]:

$$\mu_1 = 1 - \frac{F\omega_{mp}^2}{\omega^2 - \omega_{mp}^2 - i\Gamma(\omega)\omega}, \qquad (1)$$

where $\Gamma(\omega) = [\omega^2/(\omega_r+\omega_m)+\omega_r+\omega_m]\alpha$; $\omega_{mp} = \sqrt{\omega_r(\omega_r + \omega_m)}$; $\omega_m = 4\pi M_s \gamma$; $\omega_r = \gamma H$; $\alpha$ is the damping coefficient of ferromagnetic precession; $\gamma$ is the gyromagnetic ratio; $F = \omega_m/\omega_r$; $\omega_m$ and $\omega_r$ are the characteristic frequency and ferromagnetic resonance frequency of the ferrite, respectively; $M_s$ is the saturation magnetization caused by the applied magnetic field; and $H$ is the applied magnetic field. According to Eq. (1), the permeability of the ferrite can be tuned by adjusting the applied magnetic field. Dielectric metamaterial unit cells support an electric and magnetic dipole response attributable to Mie resonances. Proper control of the electromagnetic parameters of the unit cell allows for a modulation over the effective permittivity and permeability of the entire metamaterial. The effective permittivity of a standard cylindrical dielectric resonator can be expressed as follows[19]:

$$\varepsilon_{eff,a} \approx A\left[1 + BJ_1\left(\sqrt{\mu_1\varepsilon_1}\,kr\right)\right], \qquad (2)$$

where $\varepsilon_1$ and $\mu_1$ are the permittivity and permeability of the dielectric cylinder, respectively; $k$ is the wavenumber; $J_1$ is the Bessel function of order 1; $r$ is the radius of the cylinder; $A$ and $B$ denote the size factor and the item associating with electromagnetic parameters of the cylinder, which are defined in Refs. [19] and [20]. From Eq. (2), we observe that the effective permittivity $\varepsilon_{eff}$ is influenced by the permeability $\mu_1$ and the permittivity $\varepsilon_1$ of the dielectric rod. For the metamaterial composed of ferrite unit cells, the effective permittivity $\varepsilon_{eff}$ can be estimated by [19]

$$\varepsilon_{eff} \approx \int \varepsilon_{eff,a} f(a)\,da, \qquad (3)$$

where $f(a)$ is the cell fraction function, which can be treated as a probability density distribution function of a unit cell with cell size $a$. From Eqs. (1) - (3), it can be observed that the permittivity of the metamaterial can be affected by the permeability of the ferrite rod, which can be tuned by adjusting the applied magnetic field.

Figure 2a shows the simulated scattering spectra for the unit cell of the metamaterial under a series of applied magnetic fields *H*. A transmission dip appears at 11.24 GHz in the absence of an applied magnetic field. When a magnetic field of $H = 100$ Oe is applied, a transmission dip occurs at 11.13 GHz. When *H* is increased from 100 Oe to 1000 Oe, the resonance frequency of the transmission dip increases to 11.51 GHz and exhibits magnetically tunable behavior. The effective permittivity of the unit cell of the metamaterial under the same series of applied magnetic fields *H* was extracted from the simulated scattering parameters using a well-developed retrieval algorithm[22-24]. As shown in Fig. 2b, in all cases, remarkable frequency dispersion occurs in the range of 11-12 GHz. The resonance frequency increases as *H* increases, consistent with the behavior observed in Fig. 2a. In addition, the peak value of the effective permittivity increases until it reaches a maximum and then decreases as *H* increases further. Figure 2c shows the dependence of the effective permittivity of the unit cell of the metamaterial on the magnetic field *H* at 11.296 GHz. The real part of the effective permittivity increases until it reaches a maximum (approximately 229) at $H = 500$ Oe and then decreases as *H* increases further. The imaginary part of the effective permittivity exhibits similar behavior.

To clarify the underlying physics of the resonance modes with or without applied magnetic field, we simulated the electromagnetic field distribution for the ferrite rod by using CST microwave studio. Figure 3a and 3b show the electric field distribution in *yz*-plane and magnetic field distribution in *xy*-plane for the ferrite rod with applied magnetic field $H = 0$ at 11.24 GHz, respectively. It can be seen that the induced circulation of displacement currents appears in the ferrite rod (Fig. 3a), which leads to a nonzero magnetic dipole momentum, resulting in a large magnetic field along *x* axis (Fig. 3b), demonstrating a magnetic resonance characteristic. Figure 3c and 3d show the electric field distribution in *yz*-plane and magnetic field distribution in *xy*-plane for the ferrite rod with applied magnetic field *H* = 500 Oe at 11.296 GHz, respectively. One can see that the electromagnetic field distribution for the ferrite rod with the applied magnetic field is much different from that without applied magnetic field, which is caused by the magnetization of the ferrite. The linearly polarized displacement currents are excited with a resonant pattern similar to electric dipole characteristic (Fig. 3c). Concomitantly, the magnetic field distribution in Fig. 3d shows a vortical pattern supporting the interpretation of an electric dipole-related resonance. On the basis of this analysis of the electromagnetic response of a ferrite rod with the identification of the magnetic and electric Mie-type resonances, we realized that the electric Mie-type resonance instead of magnetic resonance moves to the frequency region of 11 GHz - 12 GHz after the magnetization of the ferrite rod. It is known that resonant permittivity will generally lead to a dispersion curve. Therefore, we can obtain the effective permittivity near the electric resonance mode by analyzing the dispersion properties of the ferrite rods.

To confirm the results of the above simulations, experimental investigations of the electromagnetic properties of this metamaterial were conducted. The microwave measurement system and a photograph of the metamaterial are shown in Fig. 4a. Figure 4b presents the experimental transmission spectra for the metamaterial under a series of applied magnetic fields *H*. First, in all cases, a transmission dip occurs in the transmission spectrum. Second, when *H* is increased from 100 Oe to 1000 Oe, the resonance frequency of the transmission dip increases from 11.07 GHz to 11.52 GHz, similar to the results presented in Fig. 2a.

The real parts of the effective permittivities retrieved from the experimental scattering parameters under the same series of applied magnetic fields *H* are depicted in Fig. 4c. The results reveal remarkable frequency dispersion in the range of 11-12 GHz in all cases. The resonance frequency increases as *H* increases. In addition, the peak value of the effective permittivity increases until it reaches a maximum and then decreases as *H* increases further; thus, magnetically tunable behavior is demonstrated. The dependence of the effective permittivity of the metamaterial on the magnetic field *H* at a series of frequencies *f* is presented in Fig. 4d. In all cases, the real part of the effective permittivity reaches a maximum value at a certain *H*, confirming the results presented in Fig. 2b. At 11.284 GHz, the real part of the effective permittivity exhibits a relatively high value (approximately 201) at $H = 500$ Oe. From the analysis presented above, it is evident that the behavior of the experimental data is in good agreement with that of the simulated data. In addition, the metamaterial exhibits a giant MD effect, $[\varepsilon'(H) - \varepsilon'(0)]/\varepsilon'(0) = 15000\%$ at 11.284 GHz. The inset provides a close-up of the plot of the effective permittivity vs. the magnetic field. It is clear that the effective permittivity depends strongly on the magnetic field, indicating the high tunability of the dielectric properties of this metamaterial.

**Discussion**

We experimentally and numerically demonstrated a giant MD effect in a metamaterial based on ferrite rods in the GHz region attributable to the coupling of the Mie resonance and the ferromagnetic resonance, in which the effective permittivity derived from Mie resonance in ferrite unit cells is strongly dependent on the applied magnetic field, because a ferromagnetic resonance take place and dramatically change the effective permeability of the cell. The giant MD effect makes this metamaterial promising in key devices for wireless communication and radar technology.

**Materials and Methods**

**Sample fabrication.** The ferrite material chosen for this work was yttrium iron garnet (YIG) ferrite. Commercial YIG rods were cut to dimensions of $4 \times 4 \times 10.8$ mm$^3$. The saturation magnetization $4\pi M_s$, linewidth $\Delta H$, and relative permittivity $\varepsilon_r$ of the YIG rods were 1950 Gs, 10 Oe, and 15, respectively, and the same values were used in the simulations. The sample was fabricated by inserting the ferrite rods into a Teflon substrate. The distances between the rods in the *x* direction and *z* direction were 5 mm and 11.8 mm, respectively.

**Simulations.** The dimensions of the unit cell were $5 \times 5 \times 11.8$ mm$^3$. The YIG cuboid rod was modeled with dimensions of $w \times w \times h$ mm$^3$, where $w = 4$ mm and $h = 10.8$ mm. The saturation magnetization $4\pi M_s$, linewidth $\Delta H$, and relative permittivity $\varepsilon_r$ of the YIG rod were the same as those in the experiments. A plane wave was assumed for the incident electromagnetic field, with polarization conditions corresponding to an electric field along the $x$ axis and a magnetic field along the $y$ axis. The bias magnetic field was applied in the $z$ direction. Numerical predictions of the transmission spectra were calculated using the commercial time-domain package CST Microwave Studio TM.

**Microwave measurements.** The sample was placed between two horn waveguides connected to an HP 8720ES network analyzer, as shown in Fig. 3a. The propagation of the incident electromagnetic wave was along the $y$ axis, and the electric field and magnetic field were along the $z$ and $x$ axes, respectively. The bias magnetic field provided by the electromagnets was applied in the $z$ direction.

**References**

1. Kim, K. T. & Kim, C. I. Structure and dielectrical properties of (Pb,Sr)TiO$_3$ thin films for tunable microwave device. *Thin Solid Films* **420-421**, 544-547 (2002).
2. Cole, M. W., Nothwang, W. D., Hubbard, C., Ngo, E. & Ervin, M. Low dielectric loss and enhanced tunability of Ba$_{0.6}$Sr$_{0.4}$TiO$_3$ based thin films via material compositional design and optimized film processing methods. *J. Appl. Phys.* **93**, 9218-9225, (2003).
3. Colea, M. W., Joshia, P. C., Ervina, M. H., Wooda, M. C. & Pfefferb, R. L. The influence of Mg doping on the materials properties of Ba$_{1-x}$Sr$_x$TiO$_3$ thin films for tunable device applications. *Thin Solid Films* **374**, 34-41 (2000).
4. Somiya, Y., Bhalla, A. S. & Cross, L. E. Study of (Sr,Pb)TiO$_3$ ceramics on dielectric and physical properties. *Int. J. Inorg. Mater.* **3**, 709-714 (2001).
5. Stingaciu, M., Reuvekamp, P. G., Tai, C. W., Kremer, R. K. & Johnsson, M. The magnetodielectric effect in BaTiO$_3$–SrFe$_{12}$O$_{19}$ nanocomposites. *J Mater. Chem. C* **2**, 325, (2014).
6. Chen, Y., Zhang, X.-Y., Vittoria, C. & Harris, V. G. Giant magnetodielectric effect and magnetic field tunable dielectric resonance in spinel MnZn ferrite. *Appl. Phys. Lett.* **94**, 102906, (2009).
7. Castel, V. & Brosseau, C. Magnetic field dependence of the effective permittivity in BaTiO$_3$/Ni nanocomposites observed via microwave spectroscopy. *Appl. Phys. Lett.* **92**, 233110, (2008).
8. Smith, D. R., Pendry, J. B. & Wiltshire, M. C. K. Metamaterials and Negative Refractive Index. *Science* **305**, 788-792, (2004).

9  Bi, K., Dong, G., Fu, X. & Zhou, J. Ferrite based metamaterials with thermo-tunable negative refractive index. *Appl. Phys. Lett.* **103**, 131915, (2013).
10 Schurig, D. *et al.* Metamaterial Electromagnetic Cloak at Microwave Frequencies. *Science* **314**, 977-980, (2006).
11 Chen, P.-Y., Farhat, M. & Alù, A. Bistable and Self-Tunable Negative-Index Metamaterial at Optical Frequencies. *Phys. Rev. Lett.* **106**, 105503, (2011).
12 Wu, Y., Lai, Y. & Zhang, Z.-Q. Elastic Metamaterials with Simultaneously Negative Effective Shear Modulus and Mass Density. *Phys. Rev. Lett.* **107**, 105506, (2011).
13 Zhao, Q., Zhou, J., Zhang, F. L. & Lippens, D. Mie resonance-based dielectric metamaterials. *Mater. Today* **12**, 60-69 (2009).
14 Kang, L. & Lippens, D. Mie resonance based left-handed metamaterial in the visible frequency range. *Phys. Rev. B* **83**, 195125, (2011).
15 Kuznetsov, A. I., Miroshnichenko, A. E., Fu, Y. H., Zhang, J. & Luk'yanchuk, B. Magnetic light. *Sci. rep.* **2**, 492, (2012).
16 Zhang, F. L., Kang, L., Zhao, Q., Zhou, J. & Lippens, D. Magnetic and electric coupling effects of dielectric metamaterial. *New J. Phys.* **14**, 033031, (2012).
17 Zhao, Q. *et al.* Experimental Demonstration of Isotropic Negative Permeability in a Three-Dimensional Dielectric Composite. *Phys. Rev. Lett.* **101**, 027402, (2008).
18 Zhao, Q. *et al.* Isotropic negative permeability composite based on Mie resonance of the BST-MgO dielectric medium. *Chin. Sci. Bull.* **53**, 3272-3276, (2008).
19 Peng, L. *et al.* Experimental Observation of Left-Handed Behavior in an Array of Standard Dielectric Resonators. *Phys. Rev. Lett.* **98**, (2007).
20 Tsang, L., Kong, J. A. & Ding, K. H. *Scattering of Electromagnetic Wave*. Vol. 1 41 (Wiley, 2000).
21 Zhao, H., Zhou, J., Kang, L. & Zhao, Q. Tunable two-dimensional left-handed material consisting of ferrite rods and metallic wires. *Opt. Express* **17**, 13373-13380 (2009).
22 Smith, D. R., Schultz, S., Markoš, P. & Soukoulis, C. M. Determination of effective permittivity and permeability of metamaterials from reflection and transmission coefficients. *Phys. Rev. B* **65**, 195104 (2002).
23 Croënne, C., Fabre, B., Gaillot, D., Vanbésien, O. & Lippens, D. Bloch impedance in negative index photonic crystals *Phys. Rev. B* **77**, 125333 (2008).
24 Chen, X., Grzegorczyk, T. M., Wu, B., Pacheco, J. & Kong, J. A. Robust method to retrieve the constitutive effective parameters of metamaterials. *Phys. Rev. E* **70**, 016608 (2004).

## Acknowledgments

This work was supported by the National High Technology Research and Development Program of China under Grant No. 2012AA030403; the National Natural Science Foundation of China under Grant Nos. 51402163, 61376018, 51032003, 11274198, 51102148 and 51221291; and the China Postdoctoral Research Foundation under Grant Nos. 2013M530042 and 2014T70075.

## Author contributions

J.Z. conceived and designed the experiments. K.B. and X.M.L. performed the experiments and the numerical calculations. K.B. and J.Z. wrote the paper. All authors contributed to scientific discussions and the critical revision of the article.

## Additional information
Competing financial interests: The authors declare no competing financial interests.

**Figures**

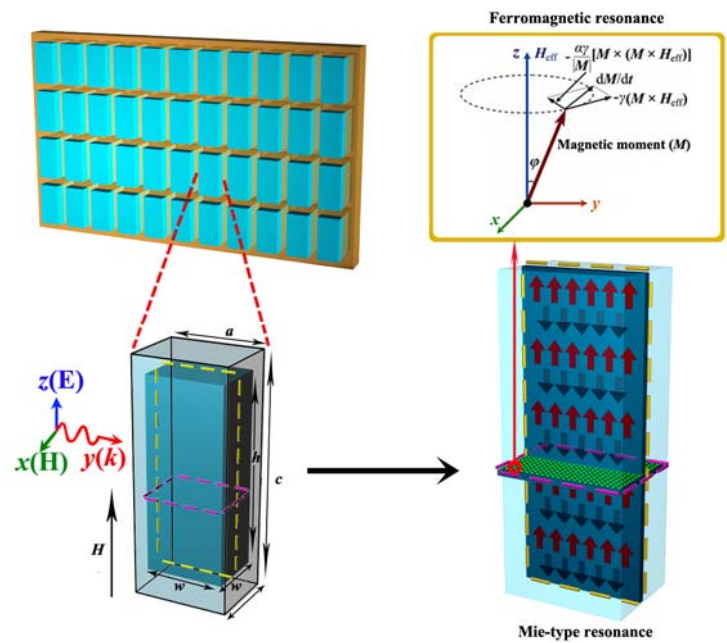

**Figure 1** | **Schematic diagram of the ferrite-based metamaterial illustrating the effect of the ferromagnetic resonance on the electric Mie-type resonance.** The length of the ferrite rod is parallel to the *z* axis. The propagation of the incident electromagnetic wave is along the *y* axis, and the electric field and magnetic field are along the *z* and *x* axes, respectively. The bias magnetic field is applied in the *z* direction.

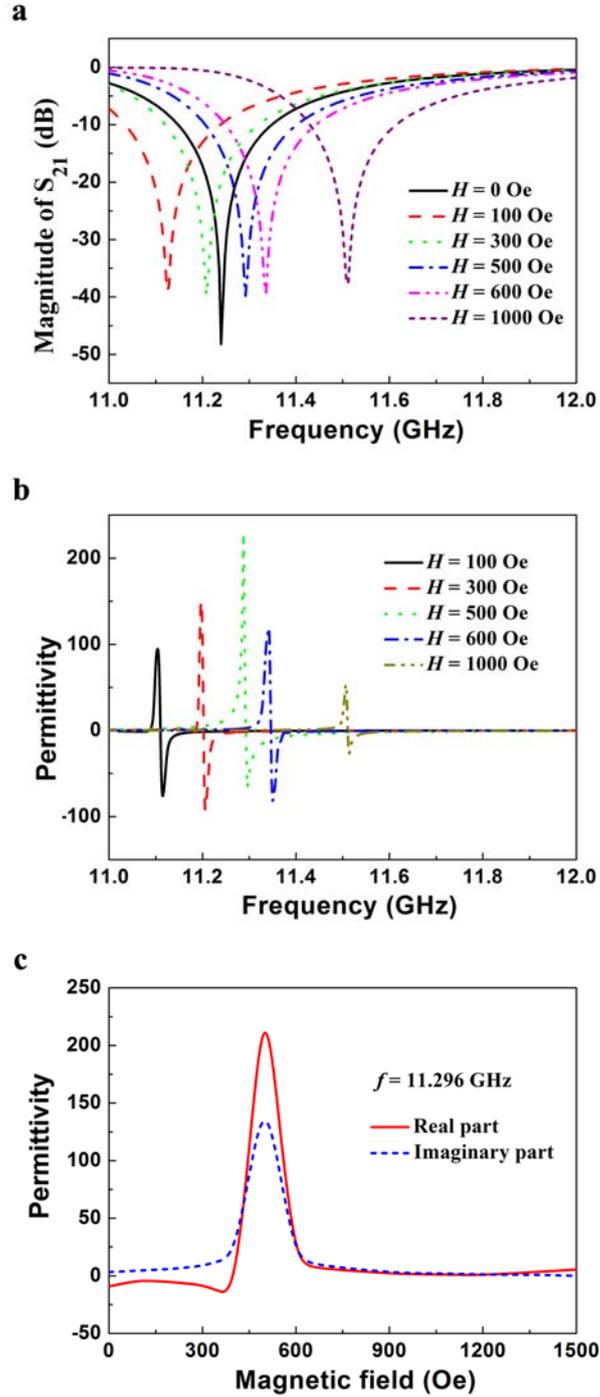

**Figure 2 | Simulated transmission spectra.** (**a**) Simulated transmission spectra for the unit cell of the metamaterial under a series of applied magnetic fields *H*. (**b**) Real parts of the effective permittivities retrieved from the simulated scattering parameters under a series of applied magnetic fields *H*. (**c**) Magnetic-field dependence of the effective permittivity of the unit cell of the metamaterial at 11.296 GHz.

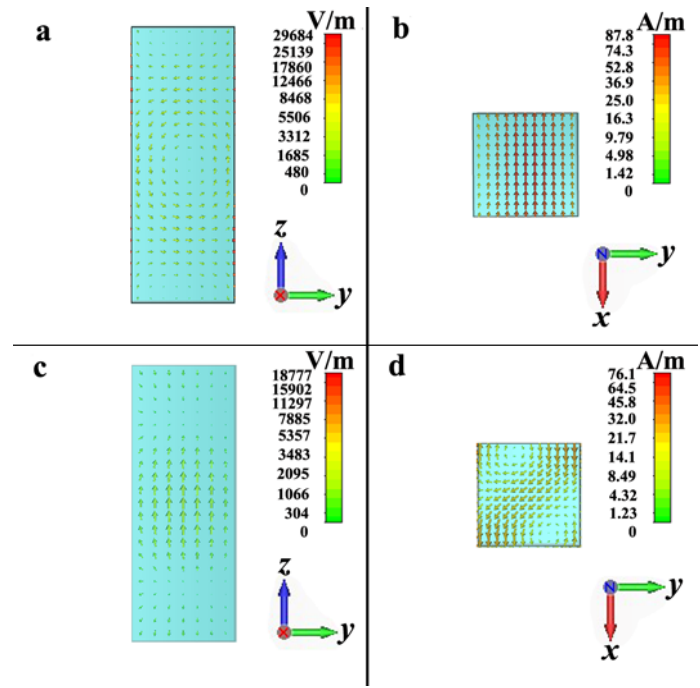

**Figure 3** | **Electric and magnetic field distributions.** Simulated (**a**) electric field distribution in *yz*-plane and (**b**) magnetic field distribution in *xy*-plane for the ferrite rod with applied magnetic field $H = 0$ at 11.24 GHz; Simulated (**c**) electric field distribution in *yz*-plane and (**d**) magnetic field distribution in *xy*-plane for the ferrite rod with applied magnetic field $H = 500$ Oe at 11.296 GHz.

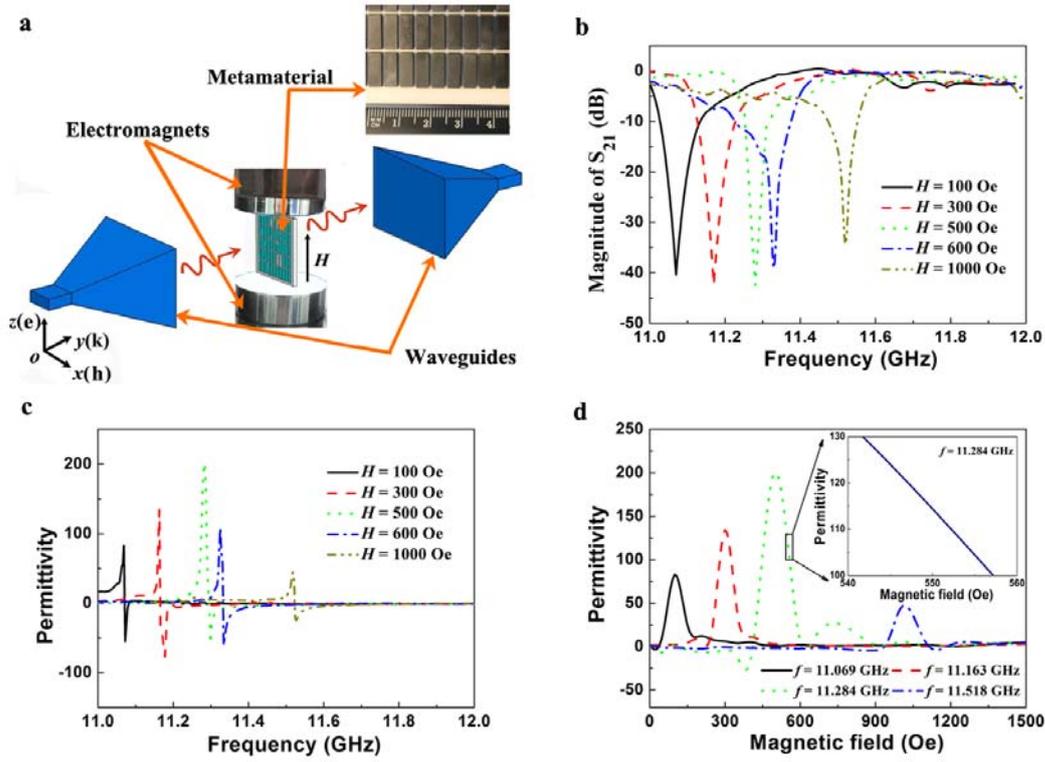

**Figure 4 | Experiment demonstrating the magnetodielectric effect.** (**a**) Diagram of the experimental setup. The sample is placed between two horn waveguides. The propagation of the incident electromagnetic wave is along the *y* axis, and the electric field and magnetic field are along the *z* and *x* axes, respectively. The bias magnetic field is generated by an electromagnet in the *z* direction. The inset shows a photograph of the metamaterial. (**b**) Experimental transmission spectra for the metamaterial under a series of applied magnetic fields *H*. (**c**) Real parts of the effective permittivities retrieved from the experimental scattering parameters under a series of applied magnetic fields *H*. (**d**) Magnetic-field dependence of the effective permittivity of the metamaterial at a series of frequency *f*. The inset provides a close-up of the plot of the effective permittivity vs. the magnetic field that illustrates the strong dependence of the permittivity on the magnetic field.